\documentclass[%
 reprint,
superscriptaddress,
nofootinbib,
 amsmath,amssymb,
 aps,
]{revtex4-1}
\usepackage{graphicx}
\usepackage{dcolumn}
\usepackage{bm}

\graphicspath{ {images/} }

\begin{document}

\preprint{APS/123-QED}

\title{A Crossover phase transition based on the phenomenology of hadronic matter and quark matter}

\author{M. Shahrbaf}
\affiliation{%
Institute for Theoretical Physics, University of Wroclaw, Max Born Pl. 9, 50-204, Wroclaw, Poland
}%

\date{\today}

\begin{abstract}
The phase transition calculations are utilized for an in-depth understanding of the thermodynamics of the deconfinement transition to perform the best analysis of the QCD phase diagram. The phenomenological justifications for a mathematical approach to constructing the phase transition are found based on the properties of hadronic matter at low densities as well as quark matter at high densities. We modify both the quark matter equation of state by confining effects at low densities and the hadronic matter equation of state by excluded volume effects at higher densities. Over the intermediate densities, the transition from the hadronic phase to the quark phase is modified by the surface tension between to phases. The results are in agreement with the observational constraints of neutron stars. 
\end{abstract}

\maketitle

\section{Introduction}
The underlying framework of this paper is carrying out a comparative study based on different ways of constructing a QCD phase transition based on the phenomenology of matter over the whole range of density. While the state of the matter at subnuclear densities is well constrained experimentally, the high-density regime (at and about twice the nuclear saturation density) is under intense scrutiny. Modern studies of compact astrophysical objects (neutron stars (NS), neutron stars in the binaries, quark and strange stars, etc.) are aimed at understanding the properties of matter under extreme conditions. Microscopic calculations of the equation of state (EoS) of dense matter profoundly affect our understanding of the origin of matter. In particular, it is required for the interpretation of an array of astronomical observations. Because of the multi-facet nature of the interiors of compact stars, we need to develop a good understanding of various phases of dense matter, ranging from ordinary nuclear and hypernuclear matter to deconfined quark matter, and their potential manifestations in the observational data. A phase transition from hadronic matter to deconfined quark matter is considered a viable solution to various puzzles in the literature. Furthermore, a significant impact of a first-order phase transition is appreciated in astrophysics \cite{Bauswein:2019skm, Bauswein:2018bma, Fischer:2016ojn, Fischer:2017lag} and the signatures of such transitions in Neutron star cores have been investigated \cite{Somasundaram:2021clp}. For a better understanding of the experimental and observational constraints of the phase transition, a QCD phase diagram within the parity doublet model is shown in Fig. 1. Some schematic features of this diagram are also depicted in the figure.
\begin{figure}[b!]
    \centering
    \includegraphics[width=0.60\textwidth]{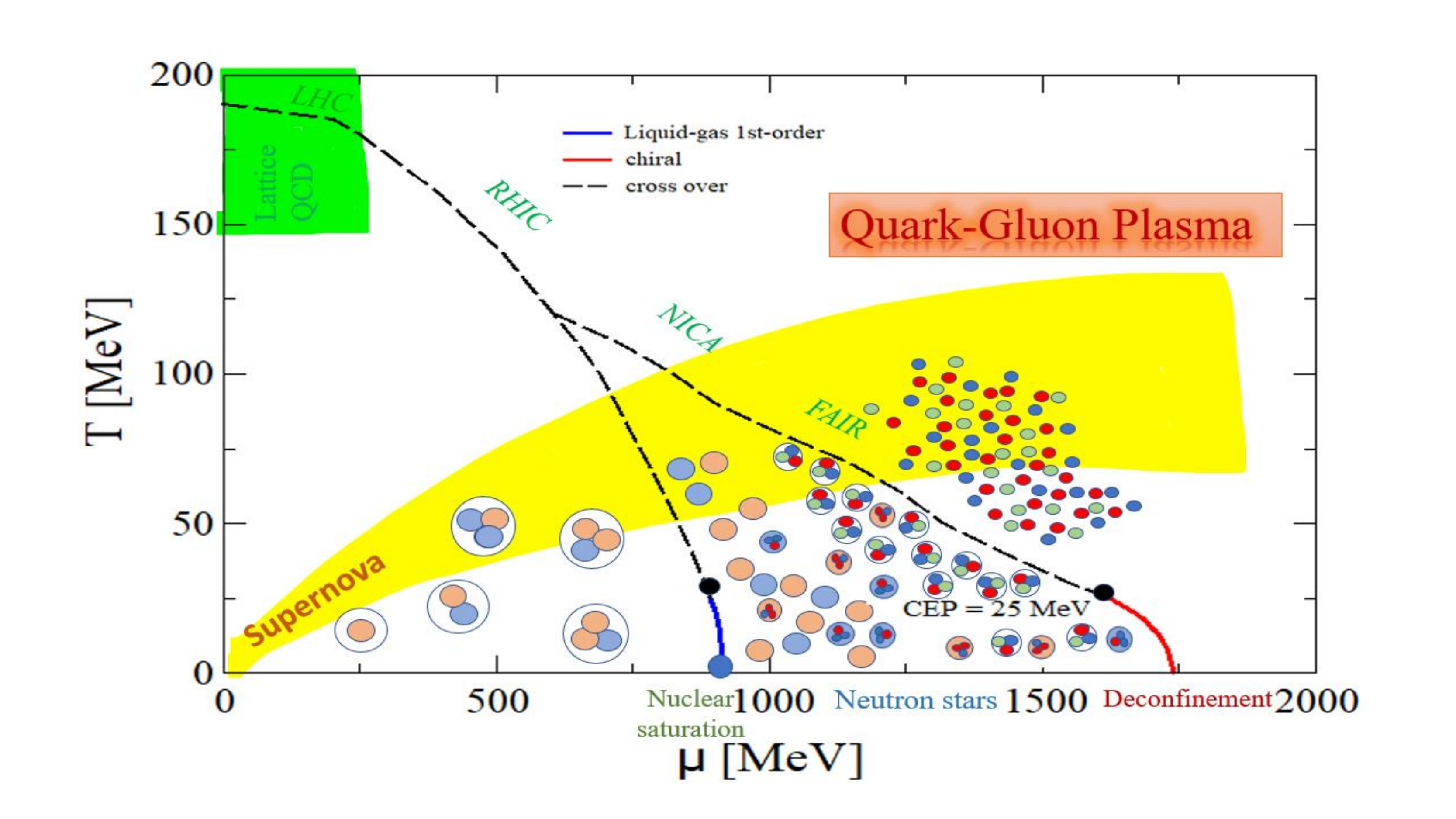}
    \caption{A schematic interpretation of the QCD phase diagram in the temperature and baryon number density plane. From lattice QCD studies at vanishing density, the phase transition is known to be a smooth crossover. The current research focus in the literature is on the search for the critical endpoint (CEP) of the first-order transition.}
    \label{fig:P-mu_2}
\end{figure}
The structure of the present paper is as follows. In Sec. II we present the formalism of calculation of the equation of state (EoS) for both hadronic matter based on the density-dependent (DD) relativistic mean field (RMF) model and the quark matter. In Sec. III the crossover construction of phase transition for constructing the hybrid stars is introduced within two different approaches. The results and discussion are provided in Sec. IV. Finally, the summary and conclusion are given in Sec. V.

\section{Equation of state for hybrid stars}

\subsection{Hadronic matter EoS}

For the hadronic EoS we have employed the relativistic density-functional (RDF) model based on the DD2F parameterization \cite{Typel:2009sy} which fulfills the flow constraint from heavy-ion collision experiments \cite{Danielewicz:2002pu}. This flow
constraint version of the DD2 EoS is obtained by multiplying the density-dependent meson-nucleon couplings $\Gamma_i(n)$
of the DD2 model \cite{Typel:2009sy} for $n>n_{sat}$ with the functions
$g_i(n)$, where $i = \omega, \sigma$.
\begin{equation}
    g_\omega(n)=\frac{1+\alpha_-x^2}{1+\alpha_+x^2}=\frac{1}{g_\sigma(n)}
\end{equation}
where $x=n/n_{sat}-1$ and $\alpha_\pm=k(1\pm r)$, with the parameter values adjusted to $k=0.04$, $r=0.07$ and $s=2.25$.
We will use this model with and without excluded volume (EV) effects \cite{Typel:2016srf}. Indeed, the hadronic matter EoS could be cured with some additional excluded volume to be stiff enough at high chemical potential.

\subsection{Quark matter EoS}
In our previous work \cite{Antic:2021zbn} which we call it paper I hereafter, We have
performed a mapping between the parameters of non local Nambu-Juna-Lasinio (nlNJL) model and constant speed of sound (CSS) model in a decent range with a $\chi^2$ value that qualifies an excellent fit. The finding of this equivalence allows
to employ the simpler CSS approach instead of the covariant nlNJL model when a hybrid star EoS with color
superconducting quark matter shall be constructed. Therefore, the quark matter phase of the neutron star is described with the CSS formulation of the nlNJL model. 

The CSS model is a frequently used, simple quark matter equation of state of a simple form
\begin{equation}
\label{eq:CSS_EoS}
    P_{CSS}(\mu) = A \left( \frac{\mu}{\mu_x}\right)^{1+\beta} - B ,
\end{equation}
with three parameters: the slope parameter A, the speed of sound $c_s$ and the bag pressure B. $\beta$ is defined as $\beta = 1/c_s^2$ and $\mu_x = 1$ GeV determines the scale for chemical potential. On the other side, the nlNJL model is defined by two free parameters, the dimensionless vector and diquark coupling strengths $\eta_V$ and $\eta_D$. They are given as ratios of the vector and diquark couplings, $G_V$ and $G_D$, to the scalar coupling constant $G_S$, respectively. Their change of values results in different equations of state (EoS) with varying stiffness. In paper I, the CSS EoS formulation is fitted to the nlNJL EoS for 34 different nlNJL EoSs with parameter values in range of 0.7 $< \eta_D <$ 0.8 and 0.11 $< \eta_V <$ 0.18. The functional form that relates parameters from the two approaches to describe quark matter phase of NSs is defined in paper I and used in this work to enable the description of quark matter phase in NS consistent with nlNJL model but with simplified formulation of the CSS approach. 


\subsection{Mixed region between the hadronic and quark matter phases}

\begin{figure}[b!]
    \centering
    \includegraphics[width=\linewidth]{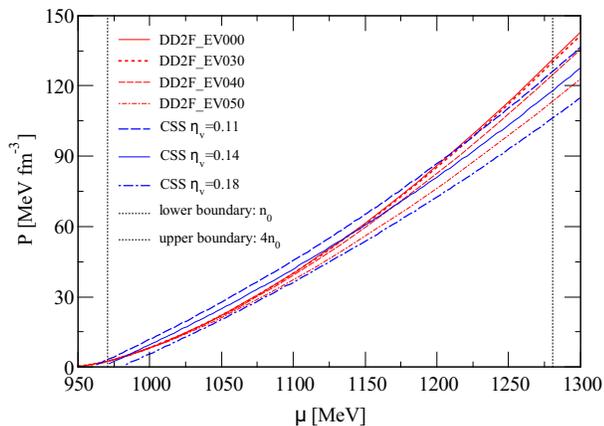}
    \caption{The DD2F model EOSs along with the excluded volume formulations of EOS for different EV parameters (EV) are shown in red while in blue we have the CSS parameterization of nlNJL model with three different values of $\eta_v$. The boundaries of saturation density and four times saturation density have been shown with black dotted lines.}
    \label{fig:P-mu_1}
\end{figure}

The region between the hadronic matter phase at below saturation densities and the expected quark matter phase ($n_B \geq 4 n_0$) is assumed to be a mixed phase of the two components with a phase transition happening at certain crossing point of the two EOSs (for HM and QM). If such a crossing exists, with HM being dominant at low densities and QM prevailing above the crossing point) we could have a Maxwell construction of the phase transition (a correct crossing point) like the low density crossing points in Fig.~\ref{fig:P-mu_1}. Otherwise, the replaced interpolation method (RIM) method must be applied in order to mimic the mixed phase between the two regions where a nonphysical crossing happens like the case at higher chemical potentials in Fig.~\ref{fig:P-mu_1} which corresponds to reconfinement \cite{Zdunik:2012dj}, keeping the HM EOS at low densities and QM EOS in high density region. The phase in between is interpolated through the mixed phase construction \cite{Shahrbaf_2020}.

Instead of using the mixed phase construction, we can mathematically model the region between the two different phases of matter through the two parabola interpolation method \cite{ayriyan2021bayesian}. The boundaries of the mixing region have to be chosen as physically reasonable ones, with  
hadronic phase ending at about saturation density ($n_H = n_0$)
and with quark matter phase onset at about four times saturation density ($n_Q = 4 n_0$). 

On top of the quark matter EoS described with CSS approach through the functional formulation with nlNJL, the additional confinement could be added in order to obtain the physical crossing point between the hadronic matter EoS and the quark matter EoS in a reasonable chemical potential range. The confinement is added as a $\mu-$dependent bag pressure,

\begin{equation}
    P(\mu) = P_{CSS}(\mu) - B_0 f_<(\mu),
\end{equation}
where $B_0$ is the strength of the bag pressure which is maximum at the beginning and decreases with increasing the chemical potential based on the definition of the switch function 
\begin{equation}
    f_<(\mu) = \frac{1}{2} \Big(1 - \frac{tanh(\mu - \mu_<)}{\Gamma_<}\Big).
\end{equation}
The switch function activates the bag pressure at low chemical potential so that the quark matter EoS includes the confinement. The range of the chemical potentials in which the bag pressure is active could be defined by the parameters $\mu_<$ and $\Gamma_<$. This behaviour is in agreement with the physics of quark matter phases, i.e., the confining effects and negative pressure at low chemical potentials for quark matter EoS is applied in this way. The effect of applying this extra bag pressure has been shown in Fig.~\ref{fig:P-mu_2}.
\begin{figure}[b!]
    \centering
    \includegraphics[width=\linewidth]{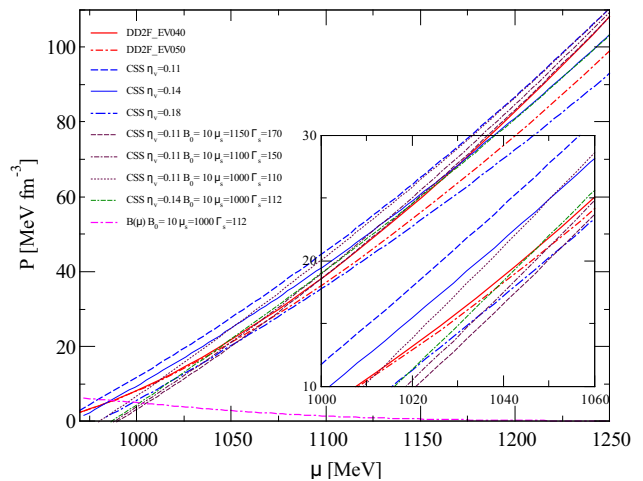}
    \caption{The DD2F-EV model EOSs for one EV parameter is shown. The CSS parameterization of nlNJL model with extra $\mu$-dependent bag pressure for different values of parameters $\mu_<$ and $\Gamma_<$ are also shown. The behaviour of bag pressure with selected parameters is plotted in magenta as well.}
    \label{fig:P-mu_2}
\end{figure}

One could see the effect of adding EV to the hadronic matter EoS as well as extra $\mu$-dependent bag pressure in curing the reconfinement phenomenon. The parameter $EV=0.05$ and the $\eta_V=0.14$, which is at the middle of the valid region in paper I, have been taken to have a deconfinement onset at a reasonable chemical potential. The values $B_0=10$ MeV fm$^{-3}$, $\mu_s=1000$ MeV and $\Gamma_s=112$ are chosen so that there will be no crossing point corresponds to reconfinement at high chemical potential and the maximum change in the slopes take places at the crossing point. It is worth mentioning that a big jump in density at the crossing point could result in an unstable branch in mass-radius curve of the NS and the twins phenomenon or even forming a black hole eventually. By choosing these parameters we could also investigate an early phase transition for which even the low mass NS could be a hybrid one which is interesting itself.

\section{Crossover construction of phase transition}

\subsection{Mixed Phase construction}
At the intermediate range of density, i.e., $n_H(\mu) < n(\mu) < n_Q(\mu)$, neither hadronic EoS nor quark matter EoS are applicable.

Since the EoS at the the intermediate range of density is not well defined, it is assumed that the EoS of both phases changes due to the finite size and Coulomb effects around the phase transition point which can be obtained using the Maxwell construction. Furthermore, the surface tension between the hadronic phase and quark phase in strongly interacting system is not well known. If the value of surface tension is infinite, a Maxwell construction can be used for phase transition while a vanishing surface tension corresponds to the Glendenning construction. 
For mixed phase transition which mimics the pasta phase, we assume a variable surface tension.\\
 The exact description of the physics of pasta phase is not straightforward because it requires the size and shape of structures as well as transitions between them to be taken in to account but it has been studied in different works by different methods \cite{Voskresensky:2002hu, Yasutake:2014oxa}.
 
In this work, we use the RIM \cite{Abgaryan:2018gqp} 
in which a simple modification of the Maxwell construction is employed instead of a full solution of pasta phase. Since the EoS of hadronic and quark matter phases are described with the relation between the pressure and chemical potential (for $T = 0$ which is applicable for the NS), $P_H(\mu)$ and $P_Q(\mu)$ respectively, the effective mixed phase EoS $P_M(\mu)$ can be described simply by an interpolated function between two phases. It requires that the interpolated pressure coincides the hadronic and quark values at lower and upper limits along with satisfying the thermodynamic constraint of positive slope of density versus chemical potential, i.e., $\frac{\partial n_M}{\partial \mu_M} = \frac{\partial^2 P_M}{\partial \mu_M^2}  > 0$, as well as the causality condition that the adiabatic speed of sound at zero frequency, $c_s^2 = \frac{\partial P}{\partial \epsilon}$, does not exceed the speed of light. A simple and reasonable function to interpolate the pressure is a polynomial function which smoothly joins the pressure curves of two phases.
This method has been developed in \cite{Ayriyan:2017tvl} and applied to the question of robustness of NS mass twins against mixed phase effects in 
\cite{Ayriyan:2017nby}. We repeat here the basic steps of its derivation following Refs.~\cite{Ayriyan:2017tvl,Ayriyan:2017nby}.

The value of the critical baryonic chemical potential $\mu_c$ for which the phases are in mechanical and chemical equilibrium is obtained from the Maxwell condition
  \begin{equation}
  P_Q(\mu_c)~=~P_H(\mu_c)~=~P_c~.
   \label{eq:mechequi}
  \end{equation}
For the pressure of the mixed phase a polynomial ansatz is considered
\begin{equation}
 P_M(\mu)~=~\sum_{q=1}^{N}\alpha_q(\mu-\mu_c)^q~+~(1+\Delta_P)P_c,
  \label{eq:RIMpressure}
\end{equation}
in which $\Delta_P$ is a free parameter which determines the pressure of mixed phase at $\mu_c$
\begin{equation}
 P_M(\mu_c)~=~P_c+\Delta_P~=~P_M~,~\Delta_P=\Delta P/P_c.
  \label{eq:addpressure}
\end{equation}

Generally in (\ref{eq:RIMpressure}), one can consider \cite{Abgaryan:2018gqp}
\begin{equation}
 N~=~2k~,~k=1,2,...
  \label{eq:nnnn}
\end{equation}
According to the Gibbs conditions for phase equilibrium (\ref{eq:mechequi}) at the matching points  $\mu_{H}$ and $\mu_{Q}$ of the mixed phase pressure with the pressure of hyperonic and quark matter EoS, the pressures and their derivative of order $k$ have to satisfy the continuity conditions
\begin{eqnarray}
P_H(\mu_{H}) &=&  P_M(\mu_{H})~,\\
P_Q(\mu_{Q}) &=&  P_M(\mu_{Q})~,\\
\frac{\partial^k}{\partial\mu^k}P_H(\mu_{H}) &=&  \frac{\partial^k}{\partial\mu^k}P_M(\mu_{H})~,\\
\frac{\partial^k}{\partial\mu^k}P_Q(\mu_{Q}) &=&  \frac{\partial^k}{\partial\mu^k}P_M(\mu_{Q})~.
\label{eq:continuity}
\end{eqnarray}
For ease of calculation, we assume that the effective mixed phase pressure could be described in the parabolic form
\begin{equation}
 P_M(\mu)~=~\alpha_2(\mu-\mu_c)^2~+~\alpha_1(\mu-\mu_c)~+~(1+\Delta_P)P_c,
  \label{eq:pppressure}
\end{equation}
for which the $\alpha_1$, $\alpha_2$ as well as $\mu_{H}$ and $\mu_{Q}$ could be obtained from the continuity conditions when $k=1$. 
%
Considering $n(\mu)=dP(\mu)/d\mu$, we numerically solve the continuity equations for baryon density and obtain $\mu_{H}$ and $\mu_{Q}$.

\begin{figure}[b!]
	\includegraphics[width=0.5\textwidth]{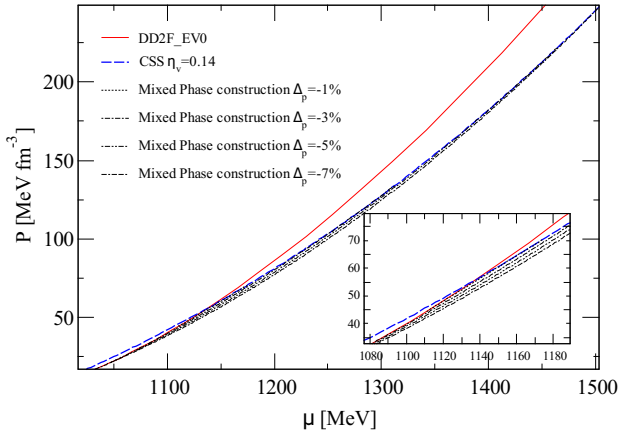}
	\caption{MP construction for a nonphysical crossing between DD2F$\_$EV0 and CSS $\eta_v=0.14$ without bag pressure with different negative values of $\Delta_p$. 
		\label{fig:MP_1}
	}
\end{figure}

\begin{figure}[t!]
	\includegraphics[width=0.5\textwidth]{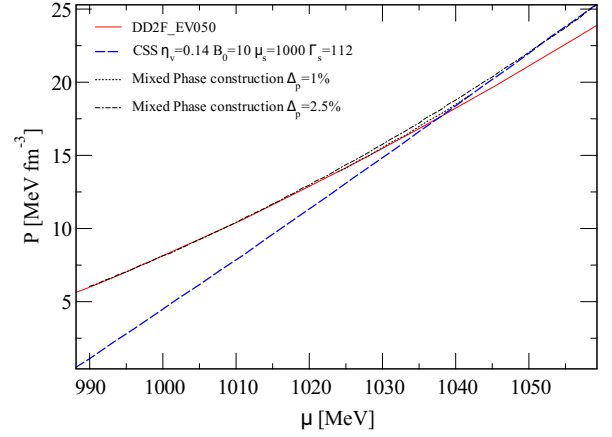}
	\caption{MP construction for a nonphysical crossing between DD2F$\_$EV050 and CSS $\eta_v=0.14$ which has been modified by a $\mu$-dependent bag pressure with given parameters for three different positive values of $\Delta_p$.
		\label{fig:MP_2}
	}
\end{figure}

It is worth mentioning that this interpolation method can be applied not only to the usual phase transition from the hadronic phase to the quark phase with 
$\Delta_P > 0$ but also to the case where applying the principle of maximum pressure to the crossing of the pressure vs. chemical potential curves for quark and hadronic matter would correspond to a nonphysical reconfinement transition from quark phase to hadronic phase with $\Delta_P < 0$. 
We have shown the results of a mixed phase (MP) construction for both physical and nonphysical crossing in Fig.~\ref{fig:MP_1} and Fig.~\ref{fig:MP_2}.

Fig.~\ref{fig:MP_1} shows the MP construction for the case that both hadronic matter EoS and quark matter EoS are the bare ones and there is a nonphysical crossing between them. As Fig.~\ref{fig:P-mu_1} shows, there is also a physical crossing for DD2F$\_$EV0 and CSS $\eta_v=0.14$ but it occurs at very low chemical potential. Therefore, we ignore it and apply a MP construction with negative value of $\Delta_p$ to the higher crossing.

The fascinating effect of adding EV to hadronic matter EoS and extra bag pressure to quark matter EoS is clear in Fig.~\ref{fig:MP_2} where not only the low density crossing has been removed but also there is only one physical crossing at reasonable chemical potential between DD2F$\_$EV050 and CSS $\eta_v=0.14$ which has been modified by a $\mu$-dependent bag pressure, $B_0=10$ MeV fm$^{-3}$, $\mu_s=1000$ MeV and $\Gamma_s=112$. Therefore, a MP construction with positive value of $\Delta_p$ could be applied.

However, the MP EoS has been constructed for some different values of $\Delta_p$, but the obtained boundaries, $\mu_H$ or $\mu_Q$, are out of the acceptable range of $n_0 < n < 4n_0$ for $\Delta_p \approx 0.3$ in both cases. Therefore, we concentrate on this value for further calculations.

Table.~\ref{tab:Boundaries} specifies the boundaries of MP construction for different models corresponding to different values of $\Delta_p$. 
\begin{table*}[t]
\caption{\label{tab:Boundaries}
 The values $n$ and $\mu$  at the hadronic and quark matter boundaries and at the critical point as well as the value of $\Delta P$ for different models. The value of $\eta_D$ is 0.75. For the models with EV, the value of EV$=0.05$ has been used. For the models with extra bag pressure, the parameters of $\mu$-dependent bag pressure are $B_0=10$ MeV fm$^{-3}$, $\mu_s=1000$ MeV and $\Gamma_s=112$.}
\begin{ruledtabular}
\begin{tabular}{cccccccccc}
$\eta_V$ & $B(\mu)$ & EV & $n_H$ & $\mu_H$ & $n_C$ & $\mu_C$ & $n_Q$ & $\mu_Q$ & $\Delta P$ \\
 & [Mev fm$^{-3}$]& & [$n_0$] & [MeV] & [$n_0$] & [MeV] & [$n_0$] & [MeV] & [Mev fm$^{-3}$]\\
\hline
 0.14 & No & No & 2.6 & 1111.0 & 2.9 & 1135.9 & 2.6 & 1174.5 & -1\% \\
 0.14 & No & No & 2.3 & 1072.6 & 2.9 & 1135.9 & 3.3 & 1313.5 & -3\% \\
 0.14 & No & No & 2.0 & 1049.3 & 2.9 & 1135.9 & 4.3 & 1463.9 & -5\% \\
0.14 & No & No & 1.9 & 1031.9 & 2.9 & 1135.9 & 5.0 & 1574.0 & -7\% \\\hline
0.14 & Yes & Yes & 1.66 & 1024.0 & 1.8 & 1038.0 & 2.33 & 1042.0 & 1\% \\
0.14 & Yes & Yes & 1.33 & 990.0 & 1.8 & 1038.0 & 2.4 & 1069.0 & 2.5\% \\
0.14 & Yes & Yes & - & - & 1.8 & 1038.0 & - & - & 4\% \\
\end{tabular}
\end{ruledtabular}
\end{table*}

\subsection{Two parabola construction}

The two parabola construction is the mathematical approach developed to describe the EoS of a mixed phase between the hadronic and quark matter phases in hybrid NSs \cite{ayriyan2021bayesian}. The pressure is given as a parabolic function of chemical potential
\begin{align}
    P_{n}(\mu)& = a_n (\mu - \mu_{H})^2 + b_n (\mu - \mu_{H}) + c_n  \\
    P_{m}(\mu) &= a_m (\mu - \mu_Q)^2 + b_m (\mu - \mu_Q) + c_m 
\end{align}
with $P_n$ defining the parabola valid for chemical potential in range $\mu_{H} < \mu < \mu_C$ and $P_m$ being a parabola for the $\mu_{C} < \mu < \mu_{Q}$ region. The boundary values of $\mu_H$ and $\mu_Q$ correspond to chemical potentials of the defined $n_{H}$ and $n_{Q}$ densities. The critical chemical potential $\mu_C$ corresponds to the Maxwell point between the hadronic and quark matter EOSs, if one exists. Otherwise, the only condition is that it has a value in range of $\mu_H < \mu_C < \mu_Q$ and it can be chosen "by hand".

At the boundaries of different phases the boundary conditions of equal values of pressure and density hold and are the following
\begin{align}
    &P_H(\mu_H) = P_n(\mu_H) \nonumber\\
    &n_H(\mu_H) = n_n(\mu_H) \nonumber\\
    &P_m(\mu_Q) = P_Q(\mu_Q) \nonumber\\
    &n_m(\mu_Q) = n_Q(\mu_Q) .\nonumber
\end{align}
The values of parabola coefficients $b_n$, $b_m$, $c_n$ and $c_m$ are therefore defined as
\begin{align}
    b_n &= n_H(\mu_H) \nonumber \\
    b_m &= n_Q(\mu_Q) \nonumber \\
    c_n &= P_H(\mu_H) \nonumber \\
    c_m &= P_Q(\mu_Q) \nonumber \\
\end{align}
while $a_n$ and $a_m$ are obtained after including the additional conditions at critical point
\begin{align}
    &P_n(\mu_C) = P_m(\mu_C) \nonumber \\
    &n_n(\mu_C) = n_m(\mu_C)
\end{align}
and are equal to
\begin{align}
    a_n &= \frac{-2 k_1 + k_2(\mu_C-\mu_Q)}{(2(\mu_C-\mu_H)(\mu_H-\mu_Q)} \nonumber \\
    a_m &= \frac{-2 k_1 + k_2(\mu_C-\mu_H)}{(2(\mu_C-\mu_Q)(\mu_H-\mu_Q)} \nonumber \\
\end{align}
with 
\begin{align}
    &\kappa_1 = n_Q(\mu_C-\mu_Q) - n_H(\mu_C-\mu_H) + (P_Q-P_H)\\ &\kappa_2= n_Q - n_H.
\end{align}




We can conclude the following:
\begin{itemize}
    \item The above equations can be used for the determination of $\mu_C$ instead of changing the $\mu_C$ value according to the crossing of the DD2 model with EV and the nlNJL(A) model with confinement
    \item The boundaries should be kept fixed according to the $\mu_H$ and $\mu_Q$ values  
\end{itemize}

\subsection{Comparison of the two approaches}

We show the comparison of the 2 parabola interpolation and the MP construction EOSs to describe nuclear matter in hybrid neutron stars. 
The figures show the comparison of the EoSs when EV and $\mu$-dependent bag pressure are included for $\Delta_p=2.5\%$. 

We calculate the $\chi^2$ value while assuming that the mixed phase values are the "expected"ones while the values of 2 parabola calculation are the "observed" ones. According to the $\chi^2$ formula
\begin{equation}
    \chi^2 = \sum_i^N \frac{(P_{\textrm{MP}}(\mu_i) - P_{\textrm{2p}}(\mu_i))^2}{\sigma_i^2} ~,
\end{equation}
where $N$ is the number of points for the chemical potential and $\sigma$ is the standard deviation of the MP construction defined as
$\sigma^2 = \frac{1}{N}\sum_i^N (P_{\textrm{MP,} i} - \overline{P_i})^2$, where $\overline{P_i}$ is the mean value of the MP model pressures, $\overline{P} = \frac{1}{N} \sum_i^N P_i$. The $P_{\textrm{MP}}(\mu_i)$ and $P_{\textrm{2p}}(\mu_i)$ are the values of pressure in mixed phase construction and for the 2 parabola one in each point of chemical potential $\mu_i$.

\section{RESULTS and discussion}
\subsection{Constraints on HM and QM EOSs: Surface tension of nuclei }
The value of $\Delta_P$ is related to the surface tension between two phases such that the vanishing $\Delta_P$ corresponds to a minimal value of the surface tension for which the transition becomes equivalent to that of the Maxwell construction in which the pressure at the critical point is constant.
The quantitative relation between $\Delta_P$ and the surface tension has been given in Ref.~\cite{Maslov:2018ghi}
for a selection of hybrid EoS cases and it shows that a value of $\Delta_P \approx 0.05-0.07$ corresponds to a vanishing surface tension and thus a Glendenning construction \cite{Glendenning:1992vb}. Therefore, it doesn't make sense to go beyond $\Delta_P=7\%$ otherwise the nature wants to show some different behaviour in contrast with our expectation.

That's why we were bound to $1\% < \Delta_P < 7\%$ and in this range, we have chosen the value of $\Delta_p$ which results in the $\mu_H$ and $\mu_Q$ in the range of $n_0 < n < 4n_0$.



\begin{figure}[h]
    \centering
    \includegraphics[width=\linewidth]{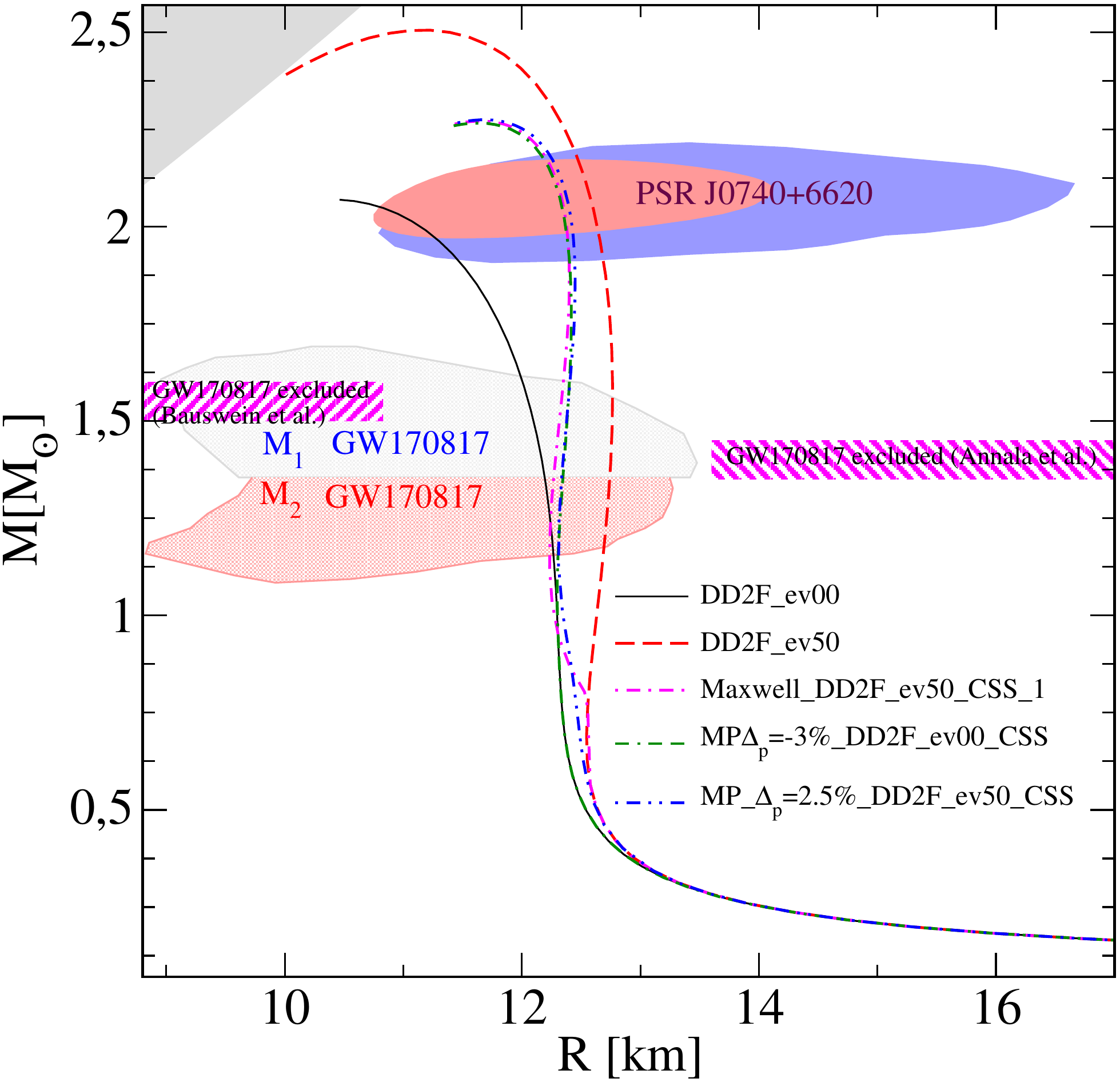}
    \caption{M-R curve for the selected parameters. The pure NSs as well as the hybrid stars with MP construction, 2p construction and Maxwell construction have been shown.}
    \label{fig:M_R}
\end{figure}


\begin{figure}[h]
    \centering
    \includegraphics[width=\linewidth] {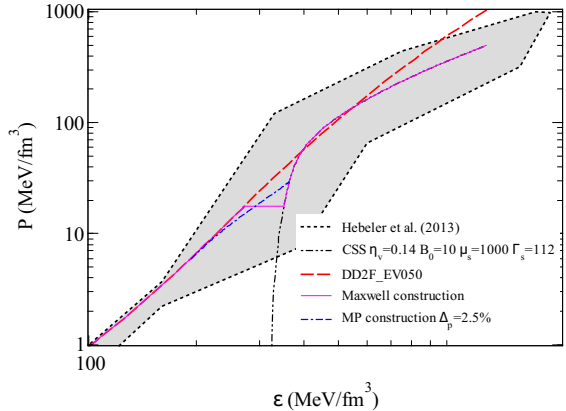}
    \caption{Pressure as a function of energy density for MP as well as 2p construction from DD2F$\_$EV050 to CSS parameterization of nlNJL model. The magenta line corresponding to Maxwell construction is also plotted to show the jump in energy density at the transition point.
The hatched region corresponds to the EoS constraint from Ref. \cite{Hebeler:2013nza}.}
    \label{fig:P-e}
\end{figure}

One of the important parameters which should be compared for MP and 2p construction is the square of the speed of sound.
\begin{equation}
    c_s^2 = \frac{\partial P}{\partial \epsilon} = \frac{1}{\beta - 1}.
\end{equation}

\begin{figure} [h]
    \centering
    \includegraphics[width=\linewidth]{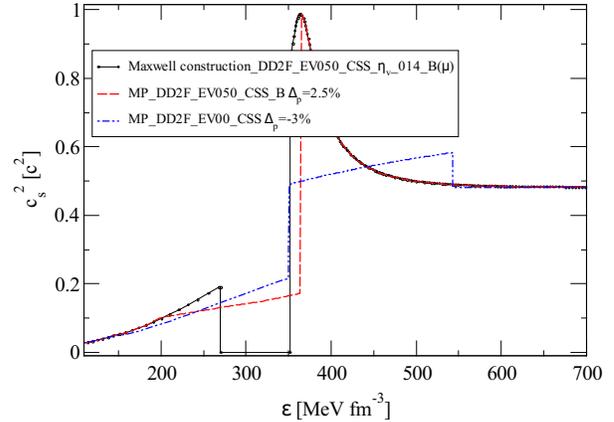}
    \caption{Speed of sound squared for the MP and 2P interpolated EOS between DD2F and CSS EoS for selected parameters.}
    \label{fig:cs2}
\end{figure}

The speed of sound squared for the interolated EOSs is given in Fig. \ref{fig:cs2}. 

As it has been mentioned in paper I, for certain nlNJL models,
the causality is violated at high values of $\epsilon$ due to the backbending of the $P-\epsilon$ curve and the maximum mass of NS could not be reached.
As a main goal of this work, using CSS parameterization we showed that not only the causality is fulfilled in the whole range of chemical potential, but also the maximum mass in agreement with the lower bound of observational values is reached and all constraints on the EoS of hybrid stars are fulfilled.

\section{CONCLUSION}
The excluded volume of the hadronic matter EoS, the bag pressure for quark matter EoS at low densities, as well as the structure of matter around the transition point which is covered in the value $\Delta_p$ in cross-over construction of the phase transition are all phenomenological justification of the two parabola construction.
\begin{acknowledgments}
This work has been supported in part by the National Science Centre (NCN) under grant No. 2019/33/B/ST9/03059 and 2018/31/B/ST2/01663.
\end{acknowledgments}

\bibliography{template}

\end{document}